\documentclass[rmp,twocolumn]{revtex4}
\usepackage{graphicx}
\usepackage{times}
\usepackage{bm}

\begin{document}
\sloppy
\DeclareBoldMathCommand{\bnabla}{\nabla}
\DeclareBoldMathCommand{\bmu}{\mu}

\title{Exploring surface interactions with atom chips}

\author{Carsten Henkel}%$^1$ and J\"org Schmiedmayer$^2$}
\affiliation{%$^1$
Institute of Physics, Potsdam University, 
Am Neuen Palais 10, 14469 Potsdam, Germany
%\\
%$^2$Physikalisches Institut, Universit\"at Heidelberg,
%Philosophenweg 12, 69120 Heidelberg, Germany
}
\email[]{carsten.henkel@physik.uni-potsdam.de}
\date{06 Dec 2005}


\begin{abstract}
%    {\bf Abstract.}
    We review the current status of the field of atom-surface
    interactions, with an emphasis on the regimes specific to atom
    chips.  Recent developments in theory and experiment are
    highlighted.  In particular, atom-surface interactions define
    physical limits for miniaturization and coherent operation.  This
    implies constraints for applications in quantum information
    processing or matter wave interferometry.  We focus on
    atom-surface interaction potentials induced by vacuum fluctuations
    (Van der Waals and Casimir-Polder forces), and on transitions
    between atomic quantum states that are induced by thermally
    excited magnetic near fields.  Open questions and current
    challenges are sketched.
%    \noindent    
\end{abstract}

\maketitle

\section{Introduction}

Atom chips provide a unique and intriguing environment to study
atom surface interactions. Two aspects can be identified: on the one 
hand, surface structures on the micron and sub-micron scale are 
crucially needed to produce electromagnetic fields that are patterned 
on this scale; this would not be possible in free space even with diffractive 
optics techniques, for example, given that the relevant wavelengths 
are much larger. On the other hand, the chip surface is a 
macroscopic, typically `hot', object whose thermal fluctuations perturb the 
ultracold atoms trapped nearby. In addition, neutral atoms are 
typically attracted to a surface by dispersion forces of the London-Van der 
Waals(-Casimir-Polder) type; this can be overcome, in some range of 
distances, by `coating' the 
surface with electromagnetic fields that provide a repulsive potential.
Ultimately, atom chip traps are metastable: the atoms are lost when 
they thermalize 
due to fluctuating near fields or tunnel through the repulsive 
potential to the surface. If a coherent operation is aimed at, 
for example in quantum information processing, then the main goal is 
to increase the corresponding time scales (trap lifetime, tunnelling time).
This often requires a compromise with the trend towards 
miniaturization because the coupling to the thermal surface is stronger
at short distance.

In this report, we review the current status of the field of 
atom-surface interactions, with an emphasis on the regimes specific to 
atom chips. Recent developments in theory and experiment are 
highlighted. Atom-surface interactions frequently define limits
in terms of a minimum height of the atom trap or a maximum time
scale for coherent operation. This implies constraints for applications
in quantum information processing or matter wave interferometry.
Finally, open questions and current challenges are sketched.

\section{Miniaturized hybrid atom-surface traps}

To exploit exceptionally long coherence times, atoms have to be
trapped in {\it vacuo\/}. Inhomogeneous electromagnetic
fields provide such traps, e.g.\ focused, off-resonant laser beams or
static magnetic quadrupole fields
\cite{Hinds99b,Grimm00b}. The feature size of these traps is
limited by the laser wavelength or the size of the magnetic coils.
Multiple, interconnected optical traps can be achieved with
micro-lenses or diffractive optics \cite{Pruvost00,Birkl02a,Birkl02b}.
Miniaturization below the micron scale requires using the
electromagnetic {\it near field\/} of a nanostructured surface. A
large variety of trapping potentials can be achieved with a network of
wires and electrodes \cite{Weinstein95,Schmiedmayer98a}, 
leading to the concept of the `atom chip'
\cite{Reichel01c,Folman02,Reichel02a,Fortagh04a}. The atoms are only indirectly
coupled to the chip surface in so far as they interact with its near
field. This approach combines versatility, massive parallelism, and
miniaturization with quantum coherence times that are still excellent.

% neutral atoms, dipole moments, 
% dipole approximation (long wavelength,
% field scale large compared to atomic size -- no chemical physics),

Atoms are neutral and couple to the field via their electric and magnetic 
dipole moments. Higher multipoles are typically neglected by making 
the long-wavelength approximation which is justified for transition 
frequencies that range from the kHz (trap vibration) via the GHz 
(hyperfine states) to the optical range (laser manipulation). More 
precisely, one also has to require that the near field gradients are 
not too large: this is the case if the typical distances between trap
and chip surface are large compared to the atomic size. In this 
`nanometer window', the chemical physics related to the transfer of 
electrons to the surface can be ignored, and electromagnetic couplings 
dominate. 

% perturbation theory sufficient
% value in quantum state (`internal dynamics'). 

It is often the case that the chip fields are small 
compared to the fields inside the atom so that perturbation theory is 
sufficient to determine the trapping potential.
A more refined 
calculation is needed when the atomic level shifts become 
comparable to the hyperfine splittings, for example in 
strongly confining magnetic traps. 
% typical atomic level structure (alkali): ground state,
% hyperfine states, Zeeman sublevels;
% excited state (fine, hyperfine, Zeeman);

For an atom at rest placed in
static magnetic or electric fields, one gets effective Hamiltonians
of the form \cite{Grimm00b,Folman02}
\begin{equation}
    {V}_{\rm mag}( {\bf r} ) = - \bmu \cdot {\bf B}( 
    {\bf r} ), \qquad
    {V}_{\rm pol}( {\bf r} ) 
= - \frac{ \alpha( 0 ) }{2 } {\bf E}^2( 
    {\bf r} )
\end{equation}
where ${\bf r}$ is the atomic position, $\bmu$ the magnetic moment,
and $\alpha( 0 )$ the static electric polarizability. Note that due to
parity conservation, atoms do not naturally have a permanent electric 
dipole moment. One gets energy level shifts by diagonalizing these 
Hamiltonians in a subspace of atomic quantum states in the ground state
manifold. 

For the alkali atoms that are typically used in atom chips, 
the ground state is split in two hyperfine components, with a 
splitting in the GHz range. The magnetic interaction $V_{\rm mag}$ 
splits each hyperfine component into magnetic sublevels that get 
separated by the Larmor energy $\hbar\omega_{\rm L} \equiv \mu_{\rm 
eff} |{\bf B}|$ where $\mu_{\rm eff}$ is the effective magnetic moment.
(This linear Zeeman effect applies when $\hbar\omega_{\rm L}$ is small 
compared to the hyperfine splitting.) It follows that the energy 
levels in a magnetic field are proportional to the magnetic quantum 
number $m$:
\begin{equation}
    {V}_{\rm mag}( {\bf r} ) = - \mu_{\rm eff} m |{\bf B}( {\bf r} )|
    .
\end{equation}
Roughly one half of the sublevels can be trapped 
around a minimum of the magnetic field (`weak field seekers' with
$\mu_{\rm eff} m < 0$) as long 
as the quantum number $m$ is conserved (the magnetic moment follows 
adiabatically). A typical example for such a magnetic trap is a 
magnetic quadrupole that can be formed by superimposing an 
homogeneous (`bias') field with the azimuthal field of a linear current 
(in a wire written onto the chip substrate). The field minimum is 
located in a line above the wire and provides a linear guide for atoms
(`side guide') \cite{Haensch98,Schmiedmayer98b}. The stability limits
of the adiabatic approximation are discussed by \cite{Sukumar97,Fortagh04a}:
the loss rate shows a typical exponential suppression as the ratio 
of Larmor splitting to trap oscillation frequency increases.

The level shifts from the DC Stark shift Hamiltonian $V_{\rm pol}$ 
are the same for all magnetic sublevels of a given hyperfine manifold 
because the electronic ground state has zero orbital angular momentum 
and the polarizability $\alpha$ reduces to a scalar. They thus
provide a state-independent trapping potential, and this has advantages 
regarding sublevel-changing collisions. The minimum energy,
however, occurs at large field strength and hence on the chip surface
(note that $\alpha(0) > 0$).
Stable electrostatic trapping thus requires additional interactions 
(magnetic, optical) to `shield' the surface \cite{Schmiedmayer98a,Folman02}.
For example, on a chip with a side guide above a wire,
additional potential minima can be created using nearby
electrodes with opposite charges \cite{Krueger03}.

% average over time for oscillating fields (microwave, optical)
% intensity gradients, phase gradients.

For time-dependent fields, one is often interested in an effective 
Hamiltonian that governs the atom center of mass motion on slower time 
scales. For optical fields, a time-average of the electric dipole 
interaction leads to (i) the so-called dipole potential, proportional to 
the light intensity and the atomic polarizability at the optical 
frequency~\cite{Grimm00b}, 
and (ii) to the radiation pressure force, proportional 
to the field's phase gradient \cite{Shimizu98,Cohen85b}. A magnetic 
analogue can be realized with microwave fields, the potential being 
proportional to the microwave power and the magnetic polarizability 
\cite{Spreeuw94}. In both cases, the atom remains in its ground state
(``dressed'' by the field) if the radiation is sufficiently far 
detuned from the optical or microwave transition.
The frequency dependence leads to a key feature of the dipole potentials:
if the time-dependent field drives the atom below (above) the resonance 
frequency, the atom is attracted to (repelled from) the field maxima,
respectively. This allows to build state-independent traps with a 
`red-detuned' optical field pattern \cite{Birkl01} or a focused laser
beam (`optical tweezer', see, e.g. \textcite{Chikkatur02a}).
Microwave pulses can also be used to imprint phase 
shifts on the atomic wave function with state-dependent signs, 
realizing one of the building blocks for qubit processing.

% electric dipole matrix elements (expression for static polarizability);
% magnetic dipole matrix elements = electron spin, nuclear spin.

% give orders of magnitude?

% general approach: first level: determine level shifts from interaction 
% with trapping fields. Slow atomic motion: adiabatic following,
% level shift = potential (corrections).
% Example: state-selective magnetic trapping potential
% Next step: perturbation from fluctuating fields (vacuum fluctuations,
% thermal). Leads to additional level shifts and induces transitions.

To summarize, the trapping potentials in atom chips are determined by 
diagonalizing the interaction of the atomic dipole moments with the chip 
fields. The eigenvalues can be used as potentials if the atoms move 
sufficiently slowly. This is generally the case for typical ultracold 
temperatures and trap depths. Magnetic fields provide potentials that 
depend on the magnetic quantum number, while electric fields give
state-independent traps. In the following sections, we apply 
perturbation theory to the eigenstates in the trapping fields to 
describe the interaction with the surface. This leads to induced level 
shifts due to both vacuum and thermal fluctuations 
(Sec.\ref{s:potentials}) and transitions induced between the atomic
quantum states (Sec.\ref{s:transitions}).

% in this context: see papers
% mixing between trapping fields, 
% stability of adiabatic approximation in presence of noise. 

In the context of miniaturized atom traps, we select as follows 
further directions of interest:
\begin{trivlist}
    \item{(a)}
    wire and field configurations that optimize
    the flexibility of trapping and guiding potentials
    \cite{Fortagh05b,Reichel05b};
    \item{(b)}
    non-metallic hybrid chips with magnetized structures 
    \cite{Vuletic98,Spreeuw05a,Jaakkola05}.
    or integrated 
    optical components (microcavities, waveguides, photonic crystals)
    \cite{Prentiss00b,Mabuchi02a,Eriksson05a}.
    This may 
    provide an enviroment with reduced electromagnetic noise (see 
    Sec.\ref{s:transitions}) and facilitate single-atom detection
    \cite{Horak03b};
    \item{(c)}
    the stability of the adiabatic approximation underlying the trapping 
    potentials in the presence of electromagnetic fluctuations (see 
    Sec.\ref{s:transitions});
    \item{(d)}
    corrections to the atomic level structure beyond the
    long-wavelength approximation.  This is particularly important for
    strongly confining potentials near nanostructures where the
    relevant length scale, given by the structure size, is much
    smaller than the field wavelength.  For strong magnetic quadrupole
    fields, see the recent papers by \textcite{Schmelcher05a,Schmelcher05b};
    \item{(e)}
    matter wave dynamics in low-dimensional guiding potentials. 
    Ultracold atoms thus provide a setting to study mesoscopic 
    transport, similar to electrons in miniaturized solid state
    structures. One advantage of atoms is that the inter-particle 
    interactions are simple to describe and can be controlled experimentally.
    For the impact of interactions on resonant tunnelling through an 
    atomic quantum dot, for example, see the recent paper by 
    \textcite{Schlagheck05}.
\end{trivlist}

A topical issue of The European Physical Journal D (\citeyear{EPJD05})
has recently been
devoted to atom chips, reporting recent advances in the 
field with contributions from both theory and experiment. Papers 
directly related to atom-surface interactions have been contributed by
\textcite{Buhmann05a,Schmelcher05b,Henkel05c,Dikovsky05,Zhang05a,Segev05};
some of them are summarized below.

% Theory: 
% 
% \cite{Buhmann05a}
% interaction with magnetodielectric surfaces
% 
% \cite{Schmelcher05b}
% steep magnetic gradient traps, atomic levels
% 
% \cite{Henkel05c}
% \cite{Dikovsky05}
% \cite{Zhang05a}
% magnetic near field noise and spin flips (calculations, optimization,
% experiments with layered chip)
% 
% \cite{Segev05}
% review on evanescent wave mirrors (atoms and molecules)

\section{Surface interaction potentials}
\label{s:potentials}

\subsection{Introduction}

% atom in front of surface: well-known problem of QED.
% Similar to interaction between two atoms.

An atom in front of a surface is a well-known, paradigmatic situation 
for cavity quantum electrodynamics (cavity QED)~\cite{Haroche92,Hinds94}. 
The 
presence of the surface introduces a distance-dependent contribution 
to the Lamb shift, the displacement of atomic energy levels due to 
virtual transitions induced by the electromagnetic vacuum (i.e., the 
field's zero point fluctuations). This energy shift is called the 
Van der Waals-London or Casimir-Polder potential, depending on the 
relative magnitude of the atom-surface distance compared to typical
atomic transition wavelengths. The same names are also attached to
the interaction between two neutral atoms that can be attributed to 
their electric dipole (quantum) fluctuations and the vacuum field.

% Van der Waals-London interaction: electric dipole
% and its (instantaneous) image. 
% attractive $1/z^3$ power law, coefficient
% involves $\langle {\bf d}^2 \rangle$, quantum-mechanical average,
% and weighted `oscillator strength' of non-retarded surface
% response $(\varepsilon( \omega ) - 1)/(\varepsilon( \omega ) + 1)$

For the alkali atoms, the electric dipole fluctuations in the ground 
state $|{\rm g}\rangle$
are dominated by virtual transitions to the first excited 
state $|{\rm e}\rangle$ (the D1 and D2 lines). The Van der Waals-London limit applies
at atom-surface distances much smaller than the corresponding wavelength,
$z \ll \lambda_{\rm eg}$. The potential is then dominated by the 
instantaneous (non-retarded) response of the surface to the electric 
dipole fluctuations and can be interpreted in terms of an image dipole.
One finds an energy shift proportional to the squared electric dipole 
matrix elements $| \langle {\rm g} | {d}_{i} | {\rm e} \rangle  |^2$, with a power 
law $1/z^3$ and a frequency average of the surface `reflectivity' in 
the non-retarded limit, weighted by the atomic polarizability
\cite{Sipe84}
\begin{eqnarray}
E_{\rm vdW}( z ) &=&
- \frac{ 1 }{ 2 \pi\varepsilon_0 (2z)^3}
\sum_{e} \left(
| \langle {\rm g} | {\bf d} | {\rm e} \rangle  |^2 
+
| \langle {\rm g} | {d}_z | {\rm e} \rangle  |^2 
\right)
\nonumber
\\
&& {}\times \,{\rm Im}
\int\limits_0^\infty\!\frac{ {\rm d}\omega }{ 2 \pi }
\frac{ \varepsilon( \omega ) - 1 }{ \varepsilon( \omega ) + 1 }
\frac{ \omega_{\rm eg} }{ \omega_{\rm eg}^2 - \omega^2 - 0 {\rm i} \omega}
,
\label{eq:vdw-shift-near-field}
\end{eqnarray}
where $\varepsilon( \omega )$ is the dielectric function of the surface
material. The summation runs over all excited states that are connected
with electric dipole transitions to the ground state; taking only
the D1/D2 transitions gives results accurate within the percent level
\cite{Klimchitskaya05b}.

% Casimir-Polder interaction: includes retardation,
% limit of distance large compared to wavelength. Image dipole
% decorrelates, reduced attraction, $1/z^4$ power law,
% coefficient involves static polarizability and 
% weighted angular average of static reflection coefficients. Explicit
% formula for metallic surface.

In the opposite regime, $\lambda_{\rm eg} \ll z$, retardation reduces 
the size of the interaction, and one gets the $1/z^4$ Casimir-Polder 
shift. Its expression involves atomic polarizability and surface 
dielectric function at zero frequency, with an angular average of 
surface reflection coefficients 
\begin{eqnarray}
    V_{\rm CP}( z ) &=& 
    - \frac{ 3 c }{
    4 \pi^2 \varepsilon_{0} (2z)^4 } 
    \sum_{e}\frac{ 
	\langle {\rm g} | d_{i} | {\rm e} \rangle  
	\langle {\rm e} | d_{j} | {\rm g} \rangle  }{
	\omega_{\rm eg} }
    \times {}
    \nonumber
    \\
    &&{}\times
    \int\limits_0^{\pi/2}\!{\rm d}\alpha\,
    \sin\alpha 
    \Big\{
    - \cos^2\alpha \,
    \Delta_{ij}\,
    r_{\rm s}( \varepsilon_{{\rm stat}} )
    \nonumber
    \\
    &&{}
    +
    \left(
    2 \sin^2\alpha \,
    \delta_{iz} \delta_{jz}
    + 
    \Delta_{ij}
    \right)
    r_{\rm p}( \varepsilon_{{\rm stat}} )
    \Big\}
    ,
    \label{eq:CP-shift}
    \\
    r_{\rm s}( \varepsilon_{{\rm stat}} ) &=&
    \frac{ 1 - \sqrt{ \varepsilon_{{\rm stat}} \cos^2 \alpha + \sin^2 
    \alpha }  }{ 1 + \sqrt{ \varepsilon_{{\rm stat}} \cos^2 \alpha + \sin^2 
    \alpha } }
    ,
    \\
    r_{\rm p}( \varepsilon_{{\rm stat}} ) &=&
    \frac{ \varepsilon_{{\rm stat}} - \sqrt{ \varepsilon_{{\rm stat}} \cos^2 \alpha + \sin^2 
    \alpha }  }{ \varepsilon_{{\rm stat}} 
    + \sqrt{ \varepsilon_{{\rm stat}} \cos^2 \alpha + \sin^2 
    \alpha } }
    .
\end{eqnarray}
This formula can be derived from the results of \textcite{Sipe84}.
We have denoted $\Delta_{ij} = \delta_{ij} - \delta_{iz} \delta_{jz}$.
For a perfect conductor and more generally for metals with a nonzero 
\textsc{dc} conductivity, $\varepsilon_{{\rm stat}} = {\rm i}\,\infty$, and
therefore $r_{\rm s} \equiv -1$ and $r_{\rm p} \equiv +1$. In that 
case, the $\alpha$-integral in Eq.(\ref{eq:CP-shift}) gives 
$\frac43 \delta_{ij}$.
For an explicit formula for all $\varepsilon_{\rm stat}$,
see \textcite{Antezza04}.

% finite temperature

Note that formulas~(\ref{eq:vdw-shift-near-field}, \ref{eq:CP-shift}) apply, 
strictly speaking, at zero temperature only. For $T > 0$, thermal 
photons must be taken into account by 
including a factor $\coth( \hbar \omega / 2 k_{B} T )$ in the 
$\omega$-integral~(\ref{eq:vdw-shift-near-field}) and by adding a term
involving the surface reflectivity at the transition frequency 
$\omega_{\rm eg}$. This enhances the contribution of low-lying 
substrate resonances (Section~\ref{s:hot-atom-surface}). The asymptotic 
expansion behind~(\ref{eq:CP-shift}) is only valid for distances
$z \ll \hbar c / k_{B} T$. For larger distances (beyond a few 
$\mu{\rm m}$ at room temperature), one finds an energy shift scaling 
like $T / z^3$ \cite{Barton97}.

% Analytical expressions restricted to simple geometries
% (planar cavity, cylinder, sphere). Difficult computational problem
% in actual atom chip geometry. Easy in non-retarded limit, but 
% restricted to sub-micron distance.

Analytical expressions for the Van der Waals-London-Casimir-Polder 
interaction are restricted to simple geometries (planar surface or 
cavity, cylinder, sphere), while one faces a difficult computational
problem in a generic atom chip geometry. This is related to the 
double integration over both frequencies and `angles of incidence'.
In the nonretarded limit (Van der Waals-London force), the surface 
response can be computed from electrostatics, using $\varepsilon( \omega )$ 
as dielectric constant. The frequency integration in 
Eq.(\ref{eq:vdw-shift-near-field}) remains to be done, however, and
is not restricted in general to a narrow range. And the result 
is of limited use for typical atom chips: with distances of a few microns, 
one is right in the transition regime to the Casimir-Polder potential where
retardation comes into play.

\subsection{Recent developments}

\subsubsection{Theory}

A large number of papers have appeared in recent years that refine 
the QED of atom-surface forces. Prominent trends include the impact 
of finite temperature (the previous formulas are restricted to $T = 
0$), and realistic models for the surface material. Non-equilibrium
situations and the consequences for Bose-Einstein condensates trapped
near a surface have been considered very recently. A selection is 
listed in the following.

% Finite temperature, lossy surface, two planes \cite{Barton97}.
Barton has given a straightforward extension of the standard field mode 
expansion to finite temperature and a planar cavity between two lossy 
surfaces \cite{Barton97}.

% Theory \cite{Boustimi02,Boustimi03} for Van der Waals interaction
% with metallic nanowire. Non-local dielectric response of material
% taken into account. Application to atom scattering, and internal
% state interferometry with an atom beam.
\textcite{Boustimi02,Boustimi03}
have considered the cylindrical geometry of a 
nanowire, taking into account the metal's non-local dielectric 
function. Applications mainly focus on 
scattering and interferometry with thermal atomic beams.
\textcite{Jhe97b} cover a similar geometry and focus on 
atom and molecule spectroscopy.

% \cite{Babb04} `real conditions' meaning accurate atomic polarizability
% and detailed model for metallic surface (conductivity, temperature).
Klimchitskaya and co-workers have used accurate data for 
frequency-dependent atomic polarizabilities and the material 
dielectric functions to demonstrate significant 
deviations in the Van der Waals-London regime compared to the 
contribution of the D1/D2 lines \cite{Babb04,Klimchitskaya05b}. 
Asymptotic formulas are given and shown to be accurate to within one percent.

% \cite{Segev05}
% review on evanescent wave mirrors (atoms and molecules)
C\^ot\'e, Segev, and co-workers have investigated 
the interaction of atoms and molecules with metallic and dielectric 
surfaces, with a particular emphasis on surfaces coated by evanescent 
light fields and on ultraslow matter wave reflection (`quantum reflection')
from the attractive interaction potential \cite{Cote97,Cote97b}. 
\textcite{Doak00} have discussed the scaling behaviour of quantum
reflection and pointed out the effect of a near-threshold
resonance state. 
Delay times in quantum reflection have been computed recently; they
allow to define an `effective mirror position' where the matter wave
is bouncing off the potential \cite{Friedrich04a,Friedrich04b}. 
For a review, see
\textcite{Segev05}; recent experiments on quantum reflection are described 
in Section~\ref{s:atom-expts} below.

% \cite{Buhmann04,Buhmann04a} quantization scheme for dispersive and absorbing
% surface. Includes quantization of atomic center of mass motion,
% contribution of magnetic fields to force.
Buhmann and co-workers have applied a quantization scheme 
for absorbing and dispersive media to derive expressions for 
atom-surface interactions that cover also magneto-dielectrics
\cite{Buhmann04a,Buhmann04b,Buhmann05a}.  The impact of the surface is
encoded in the backscattering of the radiation of an electric point
dipole
(i.e., reflection coefficients in the Green tensor). In addition, the
atomic polarizability becomes `dressed' by the surface, leading to 
higher order corrections that do not take the simple form of a 
potential gradient.

% \cite{Antezza04} impact on collective excitations in condensate trapped
% at a few microns from an `active' (`hot'?) surface.
\textcite{Antezza04} have shown that at a distance of a few 
microns, the interaction of surface at finite temperature
with a condensate can shift the elementary excitations of the latter.
Although the surface interaction is quite weak in this range, the 
effect is experimentally accessible with high precision spectroscopy 
of the condensate motion, as reported by \textcite{Cornell04a,Harber05a}.
The accuracy reached is sufficient to detect the thermal $T / z^3$ 
regime with surface heated above room temperature. 

% Nonequilibrium force (hot substrate) \cite{Henkel02a,Buhmann04a}.
% Nanoparticles at different temperatures: repulsion possible
% \cite{Linder66,Cohen03a}.
% \cite{Persson02a} dissipative part of Van der Waals force
% for a moving (oscillating) particle above a metal. Calculation of
% dissipated power at finite temperature, for small velocity. Highlights
% the physics of non-adiabaticity; force does not derive from a potential
% in a straightforward way \cite{Buhmann04a}.
% Similar calculation in \cite{Kyasov00}: small particle above metallic
% surface. \cite{Dorofeyev02a}: general formalism between two planar
% surfaces. Retrieves many special cases. 
% Also in
% \cite{Novotny04a}, connected to friction force via fluctuation-dissipation 
% theorem.

The same theory we outlined above can also be used for sub-wavelength 
particles, as long as higher multipole remain negligible. In this 
context, non-equilibrium forces have been studied recently. 
Possibly repulsive or non-conservative forces are generic
features in this context \cite{Linder66}. 
Forces between nanoparticles at different temperatures have been 
computed by \textcite{Cohen03a}. The interaction with 
finite-temperature surfaces has been studied, paying attention to 
dissipation for a moving particle 
\cite{Persson02a}, repulsion from radiation pressure 
\cite{Henkel02a}, coupling to magneto-dielectric surfaces 
\cite{Buhmann04a}, and unusual power laws at large distance
\cite{Antezza05a}.

An example where atom-surface interactions are used as a probe for 
condensed matter physics is the work by \textcite{Horing04}.
He has considered the modification of the Van der Waals potential
when a strong magnetic field is applied to a metallic surface and 
forces its electrons on Landau orbits.

Finally, fluctuations of the Van der Waals force have been addressed
by \textcite{Ford02}, following similar work by
\textcite{Barton91a,Barton91b} on vacuum fluctuation forces between
surfaces (the Casimir force).  Force fluctuations for a nanoparticle
have been considered by \textcite{Novotny04a} 
to derive a friction force via the 
fluctuation-dissipation theorem. This is a topic
of relevance for coherent manipulations with atom chips, since 
fluctuations lead to decoherence (Section~\ref{s:decoherence}).

\subsubsection{Measurements with cold atoms}
\label{s:atom-expts}

In the following, we focus on a few aspects that have emerged 
in recent experimental 
measurements of surface-induced interaction potentials 
with cold atoms: (a) the lowering of 
the trap depth with an additional loss channel towards the surface;
(b) the roughening, at close distance, of a magnetic atom guide 
due to imperfections in the chip structures; and (c) the reflection of
ultracold atoms and condensates from the attractive surface
potential (quantum reflection).

% Trapping potential
% 
% Attractive potential distorts shape of trap and opens new
% escape channel. Above barrier ``spilling''. 

\paragraph{Trap depth.}

% Atom reflection experiments \cite{Landragin96a}.
% Used for surface-induced
% evaporation \cite{Grimm03a}. 
% 
The atom-surface attraction can be measured by balancing it with a
controllable `electromagnetic coating', as shown with atoms reflected 
from an evanescent wave mirror by 
\textcite{Landragin96a}. The same principle has been used more recently 
in surface-induced evaporation experiments by \textcite{Grimm03a}.
They control the barrier height to the surface and remove
selectively the most energetic atoms in a trap, thus reducing the
temperature down to quantum degeneracy. 
Quantitative measurements of atom-surface interactions in the retarded
limit of a few microns are difficult to perform, however, because they are
so weak. Experiments have to carefully control electric stray fields 
due to surface adsorbates \cite{Cornell04a,Harber05a}.
Lifetime measurements in traps at different heights, performed by
\textcite{Vuletic04}, have given
indications that surface interactions play a role at micrometer
heights.  Similar results have been obtained by \textcite{Reichel04a},
where a trapped atom cloud is shifted for some time closer to 
the surface. The data analysis 
is complicated by the fact that in a magnetic trap,
one has to separate the effect of the
surface potential from a distance-dependent loss mechanism that is 
due to transitions induced by thermal magnetic fields radiated by the
surface, see Section~\ref{s:transitions} and the discussion by
\textcite{Dikovsky05}.

%   Inferred from distance-dependent trap loss rate,
%   compared to predictions for loss due to time-dependent magnetic fields
%   (see below) \cite{Vuletic04}. New distance range: retarded with respect
%   to electric dipole transition. 
%   Actually difficult to extract quantitatively: model for trapping volume,
%   accurate subtraction of distance-dependent trap loss rate
%   (see \cite{Dikovsky05}).

\paragraph{Trap roughening.}

% Roughness of static magnetic fields

% Wires on chip are imperfect (fabrication). Roughness of side walls
% gives magnetic field component along the wire. small fraction, but
% trap minimum defined by cancellation between large fields (wire and
% bias). Gives inhomogeneous potential along guide axis, detected by
% inhomogeneous atomic density. `fragmentation'

% Experiment \cite{Zimmermann02b}, 
% longitudinally inhomogeneous potential also in \cite{Ketterle02a}

% Theory: \cite{Lukin04a}

The wires on an atom chip substrate show imperfections on a scale of
nanometers to microns, depending on the fabrication process. 
The process called
`electroplating' creates wires with stronger inhomogeneities and
larger side wall roughness compared to electron beam nanolithography.
The electric current in a rough wire has nonzero components perpendicular
to the wire axis that create spatially inhomogeneous, non-azimuthal 
magnetic fields. These fields lead to a rough trapping potential when
combined with a large uniform bias field. Since the trap minimum is
defined by the cancellation of two large fields (wire and bias),
a very small angular deviation of the wire field ($10^{-4}\,$rad)
is sufficient to produce an inhomogeneous potential that fragments
a cold atom cloud as the trap approaches the wire. This 
phenomenon has been observed by several groups 
\cite{Zimmermann02b,Ketterle02a} and its origin has been elucidated
in the last few years.

\textcite{Zimmermann02b} have performed
careful symmetry checks to demonstrate that the fragmentation potential 
arises from a small, current-induced magnetic field along the wire axis.
A statistical theory for rough wire edges has been worked out
by \textcite{Lukin04a} to explain the quasi-periodicity
observed in fragmented clouds. Est\`eve and co-workers
have measured the anomalous magnetic field component from the
density distribution of a cold atom cloud and have linked that
field to the current flowing along rough wire edges, as observed with 
electron microscopy \cite{Esteve04}. The overall picture has emerged
that small-scale edge roughness gives magnetic fields that decay faster
with distance so that a broad maximum emerges in the spatial frequency
spectrum of the trapping potential, defining 
a dominant roughness length scale. 
\textcite{Wildermuth05} have shown that Bose-condensed
clouds provide a magnetic field sensor with unique performances in 
terms of field sensitivity and spatial resolution.

From the viewpoint of coherent atom processing, a rough guiding
potential can be avoided with
high-quality fabrication technology, for example, by writing the chip 
structures with a direct electron beam.
But a rough guide can also be turned into a physically 
interesting system, by realizing disorder typical for solid-state 
environments. One may thus study paradigmatic problems related
to superfluidity, mesoscopic transport and matter wave localization.

\paragraph{Quantum reflection.}

% De Broglie wave reflected if potential varies rapidly on scale of
% wavelength. Related to breakdown of WKB approximation 
% \cite{Heller96,Cote97,Cote97b}.
%   Power law potentials, `badlands' at intermediate distance.
%   Gets more efficient for weaker surface potentials (`badland' closer to
%   surface where steeper gradients occur).

%   delay times \cite{Friedrich04a}

%   hydrogen on liquid helium film \cite{Walraven89}

%   metastable neon from Si wafer, Si relief grating \cite{Shimizu01,Shimizu02}

%   Rb condensate from Si wafer \cite{Pasquini04}
%   Comparison to ab initio theory including surface potential: 
%   reasonable agreement.

A matter wave incident on a potential that varies rapidly on the scale
of the wavelength can be reflected, with a nonzero probability, even
if the potential is attractive. The same effect occurs routinely for light
at an interface. First experiments with ultracold hydrogen atoms
have been performed by \textcite{Walraven89}.
Quantum reflection is related to the breakdown of the
WKB or short-wavelength approximation, since the latter predicts 
that the matter wave follows the corresponding classical particle
path. For the typical power-law potentials occurring in atom-surface
interactions, the WKB approximation breaks down in a limited range of 
distances: there, the potential varies rapidly on a scale
set by the incident kinetic energy 
\cite{Heller96,Cote97,Cote97b}. 

When the surface potential is weaker, 
the `bad lands', where the WKB approximation fails, approach the
surface and, what is more important, they become more pronounced. 
This has been exploited in experiments
\cite{Shimizu01,Shimizu02} to enhance the efficiency
of quantum reflection. The surface density of a silicon wafer has
been reduced by writing a relief grating, and this improved the
reflectivity for metastable neon atoms. It has been studied in 
detail to which extent the atom-surface interaction actually plays a 
role in this context \cite{Oberst05a,Oberst05b,Oberst05c}.

%   Comparison to ab initio theory including surface potential: 
%   reasonable agreement.
\textcite{Pasquini04} have demonstrated the quantum reflection
of a rubidium Bose-Einstein condensate from a silicon 
wafer. A reasonable agreement has been found with
a theory for non-interacting atoms, using a potential that interpolates
between the Van der Waals and Casimir-Polder limits. Measurements of 
this kind, using ultracold atoms, hence provide the possibility to
explore the weak long-distance tail of atom-surface 
interactions \cite{Cote97,Segev05}.

\subsubsection{Measurements with thermal atoms}
\label{s:hot-atom-surface}

\paragraph{Frequency shifts.}

% Spectroscopy at sub-micron distance.

%   Reflection of light from ultrathin vapor cell
%   \cite{Ducloy03a}. 
%   Frequency modulation around atomic resonance. Narrow
%   velocity class of atoms that graze along the cell interface
%   contributes to transient signal, free from Doppler broadening.
%   Shift of transition frequency = difference between ground state
%   and excited state shift \cite{Barton97}.
%    Dominant contribution: classical
%   dipole. Impact of resonant absorption in the cell material, and
%   thermal occupation.

The reflection of light from a thin cell filled with an atomic
vapor can be used for high-precision atomic spectroscopy.
The transient response of atoms that desorb from the cell interface
gives a narrow spectral feature that is insensitive to 
Doppler broadening. It allows to infer frequency
shifts of transitions to excited atomic levels, and hence the
difference between the corresponding energy shifts \cite{Barton97}.
The impact of resonant absorption in the cell walls and of thermal
occupied surface modes, at finite temperature, 
has been demonstrated recently by \textcite{Ducloy03a}. 
These experiment and related theoretical issues
are discussed in the review paper by \textcite{Ducloy05}. 

\paragraph{Elastic and inelastic scattering.}

%   Atomic operator for non-retarded potential has symmetry of electric 
%   quadrupole transitions. Contains matrix elements that connect
%   different fine structure states. Experiment with metastable
%   argon where inelastic scattering observed. Energy difference due
%   to fine structure splitting appears as excess kinetic energy
%   normal to the surface \cite{Boustimi01}. Large angular deviation,
%   but weak process.

At grazing incidence, even a thermal atom beam can undergo quantum 
reflection, the normal velocity component becoming extremely small. 
This has been demonstrated experimentally by 
\textcite{DeKieviet03,Droujinina}, using clever echo techniques 
with a spin-polarized He3 beam. By varying the angle of incidence, both 
non-retarded and retarded branches of the atom-surface potential could 
be probed.

The Hamiltonian of the Van der Waals potential has the symmetry
of electric quadrupole transitions. It thus has nonzero matrix elements 
between different fine structure states split by the spin-orbit 
interaction. In an experiment with a metastable argon atom beam,
the corresponding transitions have been observed via a very weak
inelastic scattered beam. The spin-orbit energy splitting
is changed into kinetic energy for the motion normal to the surface, 
leading  to a large angular deviation, as shown by 
\textcite{Boustimi01}. This weak effect could only be 
demonstrated
because of the excellent efficiency of detecting metastable atoms.

\section{Surface-induced dissipation}
\label{s:transitions}

\subsection{Introduction}

% Transition rate from second-order perturbation theory: measurement
% of field correlation spectrum at transition frequency. 

We now turn to the transitions between atomic eigenstates in the 
trapping fields that are induced by the presence of the atom chip 
surface.
The system at hand is in fact strongly out of equilibrium, the surface
being typically at a much higher temperature than the atoms. The 
useful time scale for stable trapping and coherent manipulation is 
thus set by the surface-induced transition rates. In the same way as
for the 
interaction potentials, atom chips operate in a regime where the main 
coupling is mediated by the electromagnetic field, more precisely, its 
fluctuations radiated by the surface. The dominant atomic transitions 
happen in the ground state manifold, between its hyperfine and Zeeman 
sublevels, and between the center of mass eigenstates in the trapping 
potential. For these transitions, field fluctuations are highly 
thermal, and transition rates increase linearly with temperature. 
The opposite situation applies to optical transitions to
electronically excited states that occur in laser traps. 

% 
% Fixed position: ``internal transitions'' (optical absorption, emission)
% (magnetic sublevels: magnetic trap loss)

% Origin of electro-magnetic field fluctuations

The sources of electromagnetic field fluctuations in atom chips
can be roughly categorized as follows. 
\begin{trivlist}
    \item{(a)}
    Vacuum (also called zero point) fluctuations of field modes and,
    at lower frequencies, their thermal counterpart:
    \item{(a-1)}
    One type of field modes are scattering states labelled by plane 
    wave photons incident from infinity (or from the walls of the
    experimental setup). They give rise to a spatially modulated field
    pattern upon reflection from the chip structures.
    \item{(a-2)}
    A second type of modes is provided by the radiation due to 
    polarization charges or currents inside the chip material. These 
    modes have to be taken into account separately when the 
    chip material shows absorption. At sub-wavelength distances, their
    non-propagating near fields
    actually dominate the field fluctuations and increase them orders 
    of magnitude beyond the Planck blackbody spectrum.
    \item{(b)}
    The fluctuations of the external currents and fields that provide 
    the trapping potential. In a first step, they can be calculated 
    along the same 
    lines as the static fields that form the trap, using a linearization 
procedure. This works
    provided retardation is negligible at the relevant frequencies.
    Current fluctuations at the shot noise level require a more careful
    description of electron transport through the chip wires.
\end{trivlist}
% 	
%     
% Vacuum (zero point) fluctuations of all field modes; thermal
% occupation. Types of modes: plane waves incident from infinity 
% (`photons') and polarization charges or currents inside the surface
% (`material'). Typical situation: near field, noise currents are 
% the dominant source.
% 
% Technical fluctuations of charges/currents in chip structure. 
% Shot noise. Calculation simple for narrow wires: linearize current
% dependence of magnetic traping potential. 

From a general perspective, ultracold atoms can be considered as 
sensitive detectors of the electromagnetic field close to the chip 
structures: trap equilibrium positions depend on the mean fields
(average current, electrode charge), while transitions between states 
are proportional to the local spectral density of the field 
fluctuations. 

% Optical transitions

For example, electric dipole transitions at optical frequencies
happen at a rate given by Fermi's Golden Rule that can be written in 
the form
\begin{eqnarray}
    &&\gamma_{\rm i\to f} = \frac{1}{\hbar^2}
    \sum_{k, l} \left\langle {\rm i} \right| d_{k} \left| {\rm f} \right\rangle
    \left\langle {\rm f} \right| d_{l} \left| {\rm i} \right\rangle
    S_{kl}^{E}( {\bf r}, {\bf r}; - \omega_{\rm if} )
    \label{eq:optical-decay}
    \\
    &&S_{kl}^{E}( {\bf r}, {\bf r}'; \omega )
    = \int\!{\rm d}\tau\,
    {\rm e}^{- i \omega \tau}
    \left\langle E_{k}( {\bf r}, \tau ) E_{l}( {\bf r}', 0 )
    \right\rangle.
\end{eqnarray}
Here, 
$\left| {\rm i, f} \right\rangle$ are the initial and final states and
$S_{kl}^{E}( {\bf r}, {\bf r}; -\omega_{\rm if} )$ is the local
spectral density of the electric field. It gives the strength of
field fluctuations with the transition frequency $\omega_{\rm if}$. 
%
% local spectroscopy, sensitive to electric field zero point fluctuations.
% Near field: determined by local refractive index and topography
% \cite{Henkel98b,VanLabeke99}. Detection of `local density of 
% states' \cite{Chicanne02,ColasdesFrancs02} that
% can be split into radiative and nonradiative \cite{Savasta04a},
% electric and magnetic parts \cite{Joulain03}.
It is usually identified with the local density of photon states LDOS
\cite{Barnes98,Chicanne02,ColasdesFrancs02}. Due to an asymmetry
between electric and magnetic fields in the near field, this 
identification has to be done with care, however \cite{Joulain03}.
The LDOS has also been split into radiative and nonradiative parts
\cite{Savasta04a} which is similar to the distinction made above 
between photon scattering states and material radiation. Above a 
structured substrate, the decay rate $\gamma_{\rm i\to f}( {\bf r} )$ 
is position-dependent and allows to infer local optical properties with 
a sub-wavelength spatial resolution \cite{Henkel98b,VanLabeke99}. 
This can be used in scanning near field microscopy, the advantage 
being that $\gamma_{\rm i\to f}( {\bf r} )$ is independent of 
illumination and detection conditions. Some evidence for a 
distance-dependent decay rate $\gamma_{\rm i\to f}( z )$ close to a 
dielectric has been found by \textcite{Spreeuw04}: cold 
atoms dropped onto a glass surface have been detected
via the absorption of an evanescent wave coating the surface. 

In optical traps for ultracold atoms, the decay rate~(\ref{eq:optical-decay})
determines spontaneous light scattering since the trapped atoms show
some finite admixture of the electronically excited state. Spontaneous 
scattering (${\rm e\to g}$) is detrimental to coherent atom storage
due to the random recoil momentum of the 
emitted photon. Its rate can be pushed to low levels (of the order 
of $1\,{\rm s}^{-1}$) by operating the optical trap at very large 
detuning \cite{Grimm00b}.

In the next Section~\ref{s:magnetic-loss}, we focus on transitions among the ground state 
manifold, with the atomic center of mass position assumed to be fixed. 
These transitions are due to magnetic near field fluctuations that have 
been identified recently as the main source of loss from magnetic atom chip 
traps. We then switch in Section~\ref{s:heating} to perturbations
of the quantized center of mass motion.

\subsection{Magnetic near field noise}
\label{s:magnetic-loss}

\subsubsection{Overview and theoretical methods}

In a typical magnetic trap, only a subset of Zeeman sublevels (the  
weak field seeking states). Magnetic dipole transitions to the non-trapped
sublevels can thus be observed as atom loss.
The rate of these transitions is proportional to the magnetic field
spectral density
$S_{kl}^{B}( {\bf r}, {\bf r}; - \omega_{\rm if} )$; with the 
transition frequency $\omega_{\rm if} / 2 \pi$ being in the range
$1-100\,{\rm MHz}$ for typical Larmor frequencies.
A general theory to compute the local magnetic noise spectrum
has been given by \textcite{Agarwal75a}, and \textcite{Sidles00}
have presented a calculation scheme for metallic and superconducting 
structures. 
% general theory and overview \cite{Agarwal75a,Sidles00}. 
% \cite{Scheel00b} provides consistent scheme to quantize 
% macroscopic Maxwell equations with dispersive and absorbing media.
In the context of quantum electrodynamics, Kn\"{o}ll, Welsch, and 
co-workers have developed a consistent 
scheme to quantize the macroscopic Maxwell equations with dispersive 
and absorbing media \cite{Scheel00b}. This scheme can also be used to 
cover the non-equilibrium case of a surface hotter than the vacuum 
surroundings.

\paragraph{Substrate radiation.}

% Two kinds of calculation: add up magnetic fields from noise currents
% inside the material \cite{Varpula84}. Local equilibrium: 
% fluctuation-dissipation theorem, relation to 
% ${\rm Im}\,\varepsilon( {\bf x}; \omega )$ and (local) temperature 
% $T$. 
% Technical difficulty: current not embedded in vacuum because 
% interfaces between different materials, reflection/transmission occurs.
% If neglected, easy to integrate over chip structures of arbitrary
% shape. See \cite{Henkel01a,Dikovsky05}. 

The simplest model for the thermal fluctuations inside the substrate 
is based on the assumption of local thermodynamic equilibrium. The 
fluctuation spectrum of the thermal current density is then 
proportional to the mean occupation number $(\exp{( \hbar \omega / 
k_{\rm B} T( {\bf x} ) )} - 1)^{-1} \approx k_{\rm B} T( {\bf x} ) / \hbar 
\omega$, on the one hand, and ${\rm Im}\,\varepsilon( {\bf x}; \omega )$,
the imaginary part of the (local) dielectric function, on the other.
The latter can be interpreted as the local density of states of the 
reservoir where excess field energy is absorbed. 

Starting from this model 
of the thermal currents, one computes the magnetic noise spectrum by 
adding incoherently the radiation from all volume elements occupied 
by the material \cite{Turchette00a,Sidles00,Varpula84}. 
This approach indicates that magnetic 
fluctuations increase with the volume of absorbing material. A simple
proportionality holds only in the magnetostatic regime, however, where
damping inside the material can be neglected before the field emerges 
into the vacuum above. This is typically the case when either
the distance of observation $z$ or the thickness of the metallic
chip structures are much smaller than the skin depth 
$\delta = c / (\omega \,{\rm Im}\, \sqrt{ \varepsilon } ) \approx
(\mu_{0} \sigma \omega / 2)^{-1/2}$ ($\sigma$: substrate conductivity).
One also has to take into account the boundary conditions for the 
magnetic field at the substrate-vacuum interface, otherwise magnetic 
field components parallel to the interface are over-estimated by a 
factor of order three. If this correction is neglected, the incoherent 
summation over the substrate volume provides a versatile scheme that 
can handle arbitrary chip structures \cite{Henkel01a}.

\paragraph{Equilibrium near field.}

% Other calculation: restricted to global equilibrium. Fluctuation-dissipation
% theorem directly for the fields \cite{Agarwal75a,Henkel99c}. 
% `Dissipation': ${\rm Im}\,H_{ij}( {\bf r},
% {\bf r}; \omega )$ magnetic Green tensor (field of point dipole). 
% Calculation
% requires to solve a reflection problem from the chip structure. Similar
% complexity as before, but no further integration required. Easy in 
% simple geometry where expansion in adapted coordinates is possible. 

A more accurate computation applies thermodynamic equilibrium 
statistics to the field itself, as done by 
\textcite{Agarwal75a,Henkel99c,Scheel04a}.
The noise spectrum is then
proportional to $k_{\rm B} T( {\bf x} ) / \hbar \omega$ and 
${\rm Im}\,H_{ij}( {\bf r}, {\bf r}; \omega )$, the imaginary part of 
the magnetic Green tensor (i.e., the radiation of an oscillating point 
magnetic dipole). The reflection from the chip structure is needed
to compute the Green tensor, which is a problem of similar 
complexity as in the previous approach, but no further integration is 
required. Reflection coefficients are straightforward to work out in 
simple geometries where an expansion in adapted coordinates is possible.
This approach covers also distances large compared to the skin depth,
but is restricted to a field in global thermodynamic equilibrium. 
It turns out that 
above a metallic surface where $\delta \ll \lambda$, electromagnetic 
fluctuations are dominantly magnetic if 
$z \le (\delta \lambda^3)^{1/4}$ \cite{Joulain03,Henkel04c}.  

\subsubsection{Typical scalings}

% Relevant scales: skin depth at transition frequency $\delta$, 
% distance of trap center $z$, chip geometry. 
% 
% Electromagnetic noise is dominantly magnetic above a metallic
% surface.

Model calculations for metallic layers and wires in the magnetostatic
approximation have been given by \textcite{Henkel01a,Dikovsky05}.
The relevant scales are the distance $z$ of the trap center and the
skin depth $\delta$ at the transition frequency, 
and magnetostatics applies at short distance $z \ll \delta$. There,
the magnetic noise spectrum 
is independent of frequency (white noise) and follows power laws 
as a function of distance $z$: above a half-space, $S^{B} \sim 1/z$, above a 
layer with thickness $a$, $S^{B} \sim a/z^2$, above a thin wire with 
diameter $d$, $S^{B} \sim d^2/z^3$. These power laws for the noise
spectrum are completely specified by the chip geometry.  One sees that
the noise power decreases if the amount of metallic material is
reduced.

% Intermediate distance $z \gg \delta$: power laws, colored noise.
% $1/z^4$ for half space, thick layer. 

At intermediate distances, $\lambda \gg z \gg \delta$, 
one finds again power laws 
for planar geometries, but the noise becomes colored. For a half 
space and a thick layer (compared to the skin depth), an exponent 
$1/z^4$ has been found by \textcite{Henkel99c} so that the magnetic noise 
decreases faster than in the magnetostatic regime. One reason for this
is that only thermal currents in a skin layer of the surface contribute,
another one is the very inefficient transmission of the fields through 
the metallic surface \cite{Henkel05c}. In this regime, a layer thinner
than the skin depth even enhances the magnetic noise, as discussed by
\textcite{Varpula84}.

The material properties enter via a prefactor proportional to 
the product $T \sigma$. In order of magnitude, the transition rate 
between magnetic states is at short distance
\begin{eqnarray}
    z \ll \delta: \qquad
    \gamma_{\rm i\to f} &\approx&
57 {\rm s}^{-1} (T/300\,{\rm K})
(\sigma/\sigma_{\rm Au,\,300\,K})
\nonumber
\\
&& {} \times
\sum_{kl}\frac{\mu_k\mu_l^*}{\mu_{\rm B}^2}
(Y_{kl}( {\bf r} ) \times 1\,{\rm \mu m})
\label{eq:estimate-flip-rate-nf}
\end{eqnarray}
where $\mu_{\rm B}$ is the Bohr magneton, the conductivity 
$\sigma_{\rm Au,\,300\,K}$ for gold is taken as a reference, and
$\mu_{k} = \langle {\rm i} | \mu_{k} | {\rm f} \rangle$ are the magnetic 
dipole matrix elements. The tensor $Y_{kl}( {\bf r} )$ is determined
by the chip
geometry, and scales with trap distance as mentioned above.
Noise reduction can be achieved by cooling 
with suitable alloys where $T \sigma(T)$ is not constant,
see \cite{Dikovsky05}. This approach is limited to weak wire currents
where Joule heating of the wires can be neglected; for more details,
see the papers by \textcite{Zhang05a,Schmiedmayer04c}. 
For superconducting material,
the noise is contributed by the fraction of normally conducting
electrons, see \textcite{Sidles00,Scheel05a}.

% Technical noise: wire current spectrum, $1/z^2$ power law, becomes
% dominant at distances above $\sim 100\,\mu$m. 

Finally, technical current noise gives a magnetic power spectrum with 
a $1/z^2$ power law, since the fluctuating fields vary like 
$\Delta I / z$ in the 
non-retarded regime. In atom chip experiments
by \textcite{Ketterle03a}, this noise source
dominates over thermal surface fields at distances above 
$\sim 100\,\mu$m.

\subsubsection{Recent experiments}

In the last few years, experiments with atom chips and 
microtraps have been able to 
confirm quantitatively the theory outlined above. 
% first results, yet ambiguous \cite{Zimmermann02a},
A first report of surface-induced losses from a microtrap has been 
given by \textcite{Zimmermann02a}. A 
quantitative estimate was difficult due to losses during the 
transfer of atoms into the microtrap, as pointed out by
\textcite{Ketterle03a}. 
% dominated yet by technical noise \cite{Ketterle03a}
% controlled trap distance, surface material \cite{Cornell03a},
Cornell and co-workers have achieved quantitative agreement with the 
equilibrium theory outlined above, without free parameters.
In their experiment, a quadrupole trap 
approaches surfaces of different material in a controlled 
way \cite{Cornell03a}. Typical surface-induced transition rates 
are in the range of 
$1-0.01\,{\rm s}^{-1}$ at distances of $10-100\,\mu{\rm m}$ and lead 
to a distance-dependent, additional trap loss rate.
The cross over between different power laws 
at $z \sim \delta$ could be seen, as well as a strong dependence on 
the substrate conductivity. 

% cylindrical wire, shell structure, strong impact of technical
% noise as well \cite{Jones03,Scheel04a}
A cylindrical geometry was studied by 
\textcite{Jones03}, using a wire with a shell structure. This experiment 
also showed a strong impact of technical current noise. After this
had been eliminated, good agreement with theoretical modelling was 
found, as reported by \textcite{Scheel04a}. The
increased temperature of the wire had to be taken into account.

% extremely short distance, indications for trapping potential
% distortion  \cite{Vuletic04}, see however \cite{Dikovsky05}.
At extremely short distances of down to $1\,\mu{\rm m}$, Lin and 
co-workers observed deviations with respect to the theory of near 
field-induced loss \cite{Vuletic04}.  These have been attributed to
the Casimir-Polder interaction that distorts the trapping volume and
leads to the ``spilling over'' of atoms onto the surface. See 
\textcite{Zhang05a} for a similar measurement and \textcite{Dikovsky05}
for an alternative calculation.

% Impact of layered chip structure \cite{Zhang05a}
\textcite{Zhang05a} and \textcite{Hinds05b}
have analyzed the impact of a planar, layered chip 
on the trap lifetime. It has been shown both in 
theory and experiment that if a 
highly conducting metal is deposited on a substrate with a lower 
conductivity, the substrate properties cancel out and the magnetic noise 
power essentially depends on the layer thickness and conductivity.

%\subsubsection{Noise reduction strategies}

%\ldots expand on scaling laws here
% One can also choose an optimum layer thickness 
% determined by the metal conductivity and a given confinement
% ($\sigma a$ = const.) so that one recovers a linear decrease with $T$.

\subsection{Trap heating}
\label{s:heating}

% Permanent magnetic moment: atom sensitive to magnetic fields
% also while trapped. Transitions between center-of-mass states
% in the trapping potential. Require time-dependent, inhomogeneous 
% magnetic fields.

Typical trapped atoms have a permanent magnetic moment 
that makes them sensitive to magnetic field fluctuations even 
when no transitions between Zeeman levels are involved. This is 
particularly relevant at lower frequencies where the resonances of the
center-of-mass motion are located. To couple different center-of-mass 
quantum states, spatially inhomogeneous magnetic fields 
are required. Typical resonance frequencies are given by the trap 
oscillation frequency, with values in the kHz to MHz range for the 
tightly confined geometries of atom chips. The main consequence is the
heating of an atom sample that is initially cooled down to the trap bottom.

% Dominant effect: force due to displacement of trap center.
% Determined by cancellation between wire and bias field, hence
% fluctuations `shake the trap'. Process: transition between
% harmonic oscillator states in trapping potential.
% Estimate for $0 \to 1$ transition due to wire current fluctuations
% \cite{Savard97,Gehm98,Henkel03a}
It turns out that the dominant source of trap heating are fluctuations 
in the wire currents: they change the azimuthal magnetic field 
and shift the location of the trap center, which is equivalent to a
force. Estimates for the transition rate between the two lowest
eigenstates in a one-dimensional harmonic trap have been worked out
by \textcite{Savard97,Gehm98,Henkel03a}.
%   \begin{eqnarray}
%   \Gamma_{0 \to 1} 
%   &\sim&
%   3 {\rm s}^{-1} (M / {\rm amu})
%   (\Omega / 2\pi\,100\,{\rm kHz})^3
%   \nonumber
%   \\
%   && {} \times
%   \frac{ I / {\rm A} }{ (B_{\rm b} / {\rm G} )^2 }
%   \frac{ S_I( \Omega ) }{ SN_I }
%   %
%   % = \frac{ M \Omega^3 }{ 2 \hbar } S_z( - \Omega )
%   \end{eqnarray}
%   where $\Omega/2\pi$ is the trap frequency, $B_{\rm b}$ the homogeneous 
%   bias field, and the current noise spectrum $S_I$ is normalized to shot 
%   noise level $SN_I \equiv e I$. 
This heating has been observed experimentally 
by \textcite{Reichel01b} and \textcite{Jones03}: 
current supplies with parasitic noise 
reduce the lifetime of a Bose-Einstein condensate in a surface microtrap.
Low-noise electronics and shielding resulted in significant improvements 
down to the level where intrinsic surface noise became detectable.
% observed by 
% \cite{Reichel01b,Jones03} in terms of finite condensate lifetime.
% improved with low-noise power supplies.

The heating due to thermal magnetic fields radiated by the surface 
is much smaller, the main effect being trap loss, as discussed in
Section~\ref{s:magnetic-loss}. It has been shown by
\textcite{Henkel03a,Henkel00b}
that thermal magnetic fields radiated by the surface
are spatially ``rough'' on a scale given by the trap height. The 
corresponding transition rate between trap eigenstates scales with
$(\Delta x / z)^2$ times the loss rate due to a magnetic sublevel 
change, where $\Delta x$ ($ \ll z$) is the typical size of the quantum-mechanical 
trap ground state. Heating is thus masked by trap loss, unless 
additional noise sources become dominant.

% Key quantity for thermal noise: spatial correlation function of magnetic 
% field, characterizes the `roughness' of the magnetic potential,
% correlation length. Thermal noise: correlation length $\sim$
% trap distance. Typically $z \gg a$ trap size, hence heating 
% much slower than trap loss due to spin flips. 

\subsection{Surface-induced decoherence}
\label{s:decoherence}

% Concept: decay of quantum superpositions due to coupling to environment.
% Density matrix formalism: encodes relevant quantum correlation 
% functions. Combined with average over ensemble for environment 
% fluctuations. Sum over multiple experimental runs. Ergodic average.

One of the key peculiarities of quantum mechanics are 
linear superpositions of quantum states. They are also at the heart of 
the exceptionally fast ``parallelism'' in quantum computers compared to 
classical ones. In a quantum system coupled to a  
environment, superpositions are destroyed because
the different quantum states involved get entangled with environment
states that rapidly become orthogonal \cite{Stern90,Zurek91}.  ``Open
quantum systems'', as they have been dubbed, can no longer be
described by Hilbert space vectors.
One has to use (reduced) density matrices that 
are akin to correlation functions of state vectors and that permit to retrieve
all observables pertaining to the system alone. 
% Decoherence: decay of off-diagonal elements (`coherences')
% in some basis. Superpositions
% change into statistical mixtures (in this basis).
Superpositions are 
characterized by density matrices with non-vanishing off-diagonal 
elements, also called ``coherences'', if written in the basis spanned 
by the states involved in the superposition. Decoherence is the 
process in which these elements decay to zero, leading to a density 
matrix that can be interpreted as a mixture of the basis states, that is
now weighted with classical probabilities: from being ``here and 
there'', the system has come to be ``here or there''.

\paragraph{Discrete states.}

% Typical results: transitions between basis states lead also to decay
% of coherences. Closed two-state system:
A typical result of the theory of open quantum systems is that
environment-induced transitions between basis states also suppress
coherence. For a closed two-state system $|g\rangle,\, |e\rangle$, 
for example, the off-diagonal element of the density matrix $\rho$ evolves 
like
\begin{equation}
\frac{ {\rm d} \rho_{\rm eg} }{ {\rm d}t } =
- \frac{ {\rm i} }{ \hbar } 
\langle {\rm e} | \left[ H, \, \rho \right] | {\rm g} \rangle 
- \frac12 \left( \gamma_{\rm e \to g} + \gamma_{\rm g \to 
e} \right)
\rho_{\rm eg},
\end{equation}
% Decay to more levels: add loss rates.
where we can identify a ``decoherence rate'' 
$\frac12 \left( \gamma_{\rm e \to g} + \gamma_{\rm g \to e} \right)$.
If transitions to other levels occur, that rate is further increased.

% Additional processes that also lead to decoherence (`dephasing'):
% random fluctuations of energy levels and/or transition frequencies.
Additional interactions can enhance decoherence without inducing 
transitions, a process called ``dephasing''. An example are random
fluctuations $\delta \omega_{\rm eg}(t)$ of the transition frequency
of the two-level system. They randomize the relative phase between the
components of a
superposition state, and the superposition gets lost after 
averaging. This model for an open qubit system can actually be solved 
analytically \cite{Unruh95}. 
% Example: magnetic field fluctuations along static trapping field.
% Randomizes relative phase. If correlation time $\gg$ experimental
% time scale (Markov limit), exponential decay of off-diagonal
% elements with rate 
% $\propto \Delta \mu^2 B_{nn}( {\bf r}, {\bf r}; \omega \to 0 )$.
% difference in (permanent) magnetic moment
It can be used to estimate the impact of 
magnetic field fluctuations polarized along the static trapping field
(and hence along the permanent magnetic moment of the the trapped 
atoms). If the fluctuations have a correlation time much shorter than 
other experimental time scales (e.g., for white noise), an additional
decay is found for off-diagonal elements of the density matrix 
with a rate \cite{Folman02}
\begin{equation}
    \gamma_{\rm deph} = \frac{ \Delta \mu^2 }{ 2 \hbar^2 }
S^{B}_{nn}( {\bf r}, {\bf r}; \omega \to 0 )
\label{eq:deph-rate}
\end{equation}
where $\Delta \mu$ is the difference in the magnetic moment between 
the two basis states and $S^{B}_{nn}$ is the noise spectrum for the
field component along the static trapping field. 
% In order of magnitude: comparable to spin flip processes for
% thermal magnetic noise \cite{Henkel03a}. 
Thermal surface noise gives from Eq.(\ref{eq:deph-rate}) a dephasing
rate comparable to the rate for spin flip processes, and the same 
power laws apply as above.
% A power law $1/z^4$ was found by Schroll and 
% co-workers for a \cite{Bruder03}
Wire current 
fluctuations show strongly anisotropic noise for thin wires (mainly
azimuthally polarized), and give a decoherence rate scaling like 
$1/z^4$, as shown by \textcite{Bruder03}.

% Robust coherence between Zeeman sublevels with same magnetic moment
% demonstrated in atomic clock experiment, no distance depedence 
Recent experiments by \textcite{Reichel04a} have exploited
the suppression of dephasing between states with the same magnetic 
moment ($\Delta\mu \approx 0$), in order to demonstrate an atomic clock in a
chip microtrap. The two states are in different hyperfine
manifolds and only the small nuclear magnetic moment distinguishes 
their coupling to magnetic fields. Coherent superpositions between 
them are created in a Ramsey-Bord\'e interferometer scheme. 
The dephasing between the states did not show any distance-dependent 
enhancement.

% Technical noise: low-frequency limit of
% noise spectrum. 
% can be significantly larger in case of $1/f$ noise
% (lower limit set by duration of experiment) \cite{Schoen02a}.
Technical noise in the wire currents and bias fields from the external
electronics is likely to 
give the dominant contribution to dephasing because of its increase
at low frequencies (see Eq.\ref{eq:deph-rate}), in particular if 
$1/f$ noise is present. In that limit, the noise spectrum in 
Eq.(\ref{eq:deph-rate}) has to be evaluated 
at a small cutoff frequency set by the duration of the experiment,
as shown by \textcite{Schoen02a}.

\paragraph{Integrated atom interferometry.}

% Correlation (coherence) length of atomic density matrix in position
% space. Mechanism: random scatterings with momentum transfer,
% $l_{\rm coh} \sim \hbar / \delta p$ momentum width. 
Atom chips can test the (de)coherence of matter waves by letting  
different parts of a spatially delocalized state interfere. Indeed, a 
delocalized wave in a potential, for example, is in a continuous 
superposition of position states. The degree of ``quantum-ness'' 
of this wave can be characterized from the correlation or coherence 
function $\langle \psi^*( {\bf x}, t ) \psi( {\bf x}', t ) \rangle$.
This quantity is related to the contrast of the interference from two
fictitious slits placed at ${\bf x}$ and ${\bf x}'$ \cite{Goodman}.
The range in ${\bf x} - {\bf x}'$
over which the correlation function decays to zero is called the 
correlation (or coherence) length $l_{\rm coh}$. Spatial decoherence 
is the decrease of $l_{\rm coh}$ due to coupling to an environment.
This involves in particular random scatterings that
broaden the momentum distribution, since it can be shown that
$l_{\rm coh} \sim \hbar / \Delta p$ \cite{Zurek91}.

% Rough magnetic near fields: after time comparable to spin flip rate
% (for fraction that remains trapped), coherence length reduced to
% distance from surface \cite{Henkel01a}. model with Boltzmann  
% equation.
Model calculations for cold atoms interacting with a fluctuating,
rough potential indicate that the atomic coherence length is reduced to 
the correlation length of the potential, after a time set by the power
spectrum of the fluctuations \cite{Jayannavar82,Henkel01a}.
These calculations are based on a semiclassical Boltzmann equation, 
neglecting matter wave localization. 
% condensate decoherence: \cite{Kuklov00b,Kuklov02a}, also spinor 
% condensate considered. Dephasing due to SU(2) symmetry breaking,
% decoherence due to scattering from thermal component.
\textcite{Kuklov00b,Kuklov02a} have studied the decoherence of a
trapped condensate that scatters non-condensed atoms and considered
also spinor condensates.  In the latter case, robust spin coherence
can be maintained if the atom interactions do not break a SU(2)
symmetry.
% Impact of quantum degeneracy: elementary excitations are created with
% different efficiency by random potential (suppression of long-wavelength
% excitations, phase waves, phonons). Decoherence rate reduced, coherence
% length larger \cite{Henkel04b}. 
The condensate superfluiditity which is due to atom-atom interactions,
reduces the decoherence rate, as shown by model calculations of
\textcite{Henkel04b}. This is because condensate excitations at long 
wavelengths are mainly phase waves (phonons) and are inefficiently
excited by a fluctuating potential. 

% 
% condensate interferometer \cite{Hinds01a,Reichel01d}, interactions
% make it more sensitive, and less robust with respect to potential
% fluctuations \cite{Negretti04a}.
In order to test matter wave coherence on an atom chip, an 
interferometer setup is the method of choice. Time-dependent schemes 
where a trapped cloud is split and recombined, 
have been suggested by \textcite{Berman97b,Hinds01a,Reichel01d}.
Recent experiments have been reported for elongated clouds that have
been  
split along their axial direction by Bragg diffraction \cite{Prentiss05c}
and by a magnetic grating made from a wire array \cite{Fortagh05a}. 
Interferences after splitting in the radial direction and release into 
free fall have been reported
with an optically created double well \cite{Leanhardt04a} and with
the two-wire scheme on a chip by \textcite{Prentiss05d}. These experiments
use Bose-Einstein condensates and probe the phase coherence between
the two parts via the interference pattern formed when the two parts
start to overlap in free fall.  The
relative phase can be maintained when tunnelling through the barrier
is still possible, as shown by \citet{Schumm05b}. The phase diffusion
due to quantum fluctuations of the atom number difference between the
two parts \cite{Javanainen97b,Smerzi05a} has not been conclusively
observed yet. If the barrier height exceeds 
the chemical potential of the condensate, \citet{Prentiss05d}
observed a rapid phase randomization. 

Recent calculations by \textcite{Negretti04a} indicate that
when the interferometer is closed by merging the split condensate
back into a single trap,
atom-atom interactions can enhance the sensitivity to a relative phase
shift, in particular if a controlled $\pi$ phase is imposed.  
The interference gets also more
sensitive to potential fluctuations, however.  This is related to a
dynamical instability of the odd eigenstate of the condensate in the
double-well potential, as discussed by \textcite{Stickney02b}.

\section{Conclusions}

% Recent work on miniaturized atom traps
% improved understanding
Recent work on miniaturized atom traps has improved our understanding 
of atom-surface interactions in a regime where distances are 
comparable or large compared to optical transition wavelengths, but small 
compared to hyperfine or vibrational wavelengths. The fact that cold 
atoms can be maintained at such small distance from a macroscopic, 
hot surface, highlights the weakness of the coupling via 
the electromagnetic field. At the same time, surface coupling
determines limits for miniaturization and coherent manipulation.
Strategies to circumvent these limits exist, and first 
experimental demonstrations have been reported, for example an 
integrated atom clock with ``decoherence-free'' hyperfine 
states \cite{Reichel04a}.

% experiments: mainly dissipative aspect, trap loss, heating, 
% (de)coherence. Mainly `internal' qubit (spin states). Good agreement 
% with theory in simple geometry -- physics essentially understood.
% Interaction potentials: ultracold matter wave dynamics seems good 
% diagnostics (quantum reflection, elementary excitations), not so much 
% atomic (internal level) spectroscopy. 

Experiments have probed atom-surface interactions by measuring the
motion of ultracold samples or transitions between atomic sublevels
in the trapping potentials. Trap loss due to thermally
excited magnetic
near fields is now quantitatively understood, and the origin of 
condensate fragmentation in atom guides has been traced back to 
static inhomogeneities in the microfabricated wires below.
Ultracold matter wave dynamics thus has 
emerged as a sensitive tool to probe the surface interaction 
potentials. More work is required to achieve a quantitative
understanding. But we can say that the precise control over the 
center of mass motion 
afforded by atom chips thus has opened new routes for atomic spectroscopy.

% Next step: optimized low-noise traps, perform matter wave 
% interferometry. Measure surface potentials, dephasing, impact of 
% interactions.

The next steps in atom chip development are likely to involve
optimized materials and geometries for magnetic noise reduction. 
This should enable in the near future the experimental observation 
of matter wave interference. We expect to see precision measurements 
of surface potentials, of surface-induced decoherence and dephasing,
and of the impact of atom-atom interactions.

% theory: much work on more detailed description: realistic materials, 
% geometries, 
% frequency-dependent response, finite temperature, non-equilibrium
% situation. 

On the theoretical side, much work has been invested into detailed 
models for atom chip materials and geometries, including absorption,
dispersion, and finite temperature. The unusual regimes of cavity
quantum electrodynamics provided by atom chips have been identified
and put to use in efficient calculations. 
The relevant scaling laws have been spelled out and are 
being used for future atom chip designs. Versatile computational schemes 
for magnetic near field noise calculations will be developed in the 
near future. The complexity of predicting 
the Van der Waals-Casimir-Polder 
potential in a generic geometry is likely to become 
managable using clever approximation schemes. We also anticipate an
improved understanding of mesoscopic physics
by studying condensate dynamics in small-scale structures near 
surfaces.

\begin{acknowledgments}

I acknowledge extensive discussions with my collaborators and 
friends R. Folman,
J.-J. Greffet, A. Negretti, B. Power, S. Scheel, J. Schmiedmayer,
F. Sols, M. Wilkens, Bo Zhang.
This work has been supported by the European Commission under the
Future and Emerging Technologies arm of the IST programme (contract
IST-2001-38863, ACQP) and under the Human Potential programme
(contract HPRN-CT-2002-00304, FASTNet). I am grateful for support 
from the Deutscher Akade\-mi\-scher Austauschdienst under the Procope 
programme (projects 03199RH and D/0205739).

\end{acknowledgments}

\newcommand{\bibpath}{/Users/carstenh/Biblio/Database/}
\bibliography{\bibpath journals,\bibpath bib-ac,\bibpath bib-dh,%
\bibpath bib-io,%
\bibpath bib-pz,\bibpath bib-2004,\bibpath bib-2005}

\end{document}